\documentclass{PoS}
\usepackage{bbm}
\usepackage{amsmath}
\usepackage{mathtools}
\usepackage{multirow}
\usepackage{multicol}
\usepackage{graphicx}
\usepackage{subfig}

\title{Domain Wall Fermion Simulations with the Exact One-Flavor Algorithm}

\ShortTitle{EOFA Domain Wall Fermion Simulations}

\author{\speaker{David Murphy}\\
        Department of Physics, Columbia University, New York, NY 10027, USA\\
        E-mail: \email{dmurphy@phys.columbia.edu}}

\abstract{As algorithmic developments have driven down the cost of simulating degenerate light quark flavors the relative cost of simulating single quark flavors with the Rational Hybrid Monte Carlo (RHMC) algorithm has become more expensive. TWQCD has proposed an exact one-flavor algorithm (EOFA) that allows for HMC simulations of a single quark flavor without taking a square root of the fermion determinant. We have independently implemented EOFA in the Columbia Physics System (CPS) and BAGEL Fermion Sparse-Matrix Library (BFM) for Shamir and M\"{o}bius domain wall fermions, and begun to optimize and test our implementation against RHMC. In this talk we discuss the derivation of the EOFA action, our tests of its equivalence to RHMC, and the current state of our implementation and optimization. We find, after introducing a novel preconditioning technique for the EOFA Dirac operator, that EOFA is a factor of 2.4 times faster than RHMC per molecular dynamics trajectory for the strange quark determinant on an $N_{f} = 2+1$ M\"{o}bius DWF ensemble with physical quark masses and a $24^{3} \times 64 \times 24$ volume. We expect that further improvement is possible by retuning the integrator parameters for EOFA and by continuing to optimize our code.}

\FullConference{34th Annual International Symposium on Lattice Field Theory\\
		24-30 July 2016\\
		University of Southampton, UK}

\begin{document}

\section{Introduction}

A number of recent developments in the hybrid Monte Carlo (HMC) algorithm used by the RBC/UKQCD collaboration have driven down the cost of simulating degenerate quark flavor pairs. These developments include: extensive force tuning via Hasenbush mass preconditioning \cite{Hasenbusch:2001}, the zM\"{o}bius domain wall fermion action \cite{zmobius}, reduced $L_{s}$ approximations to the light quark determinant \cite{greg_thesis}, and the use of mixed-precision methods in the conjugate gradient (CG) algorithm. 
On a recent production run of a large volume, physical quark mass $N_{f} = 2+1+1$ ensemble we observed that the strange and charm quark determinants were collectively more expensive than the light quark determinant. To address this, we have turned to exploring TWQCD's exact one-flavor algorithm (EOFA) \cite{Chen:2014hyy} as an alternative to the rational HMC (RHMC) algorithm for single quark flavors. This effort is further motivated by our $I=0$ $K \rightarrow \pi \pi$ simulations with G-parity boundary conditions, where $D^{\dagger} D$ describes four flavors and RHMC is needed for the light quark pair as well \cite{g-parity}.

\section{The Exact One-Flavor Algorithm}
\label{sec:eofa}

The exact one-flavor algorithm was introduced by TWQCD for efficient simulations of single quark flavors on GPU clusters. In a series of papers the authors first demonstrate how to construct a positive-definite pseudofermion action describing a single quark flavor for Wilson and domain wall fermions \cite{Ogawa:2009ex}. They then benchmark EOFA against RHMC, finding a 20\% speed-up and a substantially reduced memory footprint for the case of EOFA \cite{Chen:2014hyy,Chen:2014bbc}. Their construction uses block manipulations in spin space to factorize a ratio of fermion determinants as
\begin{equation}
\label{eqn:eofa_det_ratio}
\det \left[ \frac{D(m_{1})}{D(m_{2})} \right] = \frac{1}{\det \left( \mathcal{M}_{L} \right)} \cdot \frac{1}{\det \left( \mathcal{M}_{R} \right)},
\end{equation}
with $\mathcal{M}_{L}$ and $\mathcal{M}_{R}$ manifestly Hermitian and positive-definite. Contrast this with RHMC, where we instead compute 
\begin{equation}
\label{eqn:rhmc_det_ratio}
\det \left[ \frac{D(m_{1})}{D(m_{2})} \right] = \left\{ \det \left[ \frac{D^{\dagger} D(m_{1})}{D^{\dagger} D(m_{2})} \right] \right\}^{1/2}
\end{equation}
using a rational approximation to the square root. Both algorithms are equivalent in the sense that they compute the same determinant ratio, but EOFA has the advantage that it avoids the need for computing an overall fractional power of the fermion determinant. \\

TWQCD's construction in Ref.~\cite{Chen:2014hyy} begins with a factorization of the domain wall fermion Dirac operator $(D_{\rm DWF})$. For our purposes we consider the M\"{o}bius kernel, and use the following notation: $\alpha$ denotes the M\"{o}bius scale, $c = \alpha/2$ and $d = 1/2$ are the weights along the fifth dimension, $D_{w}$ is the Wilson-Dirac operator, $L_{ss'}$ contains the 5D hopping terms, and $(R_{5})_{ss'} = \delta_{s,L_{s}-1-s'}$ is the 5D reflection operator. Factoring out the terms multiplying $D_{w}$ in $D_{\rm DWF}$ results in
\begin{align}
\label{eqn:Deofa_def}
\begin{split}
\left( D_{\rm DWF} \right)_{xx',ss'} &= \big( \left( c + d \right) D_{w} + \mathbbm{1} \big)_{xx'} \delta_{ss'} + \big( \left( c - d \right) D_{w} - \mathbbm{1} \big)_{xx'} L_{ss'} \\
&= \left\{ \left( D_{w} \right)_{xx'} \delta_{ss'} + \delta_{xx'} D^{\perp}_{ss'} \right \} \cdot \bigg\{ d \left( 1 - L \right) + c \left( 1 + L \right) \bigg\}_{ss'} \\
&\equiv \left( D_{\rm EOFA} \right)_{xx',ss'} \cdot \widetilde{D}_{ss'}.
\end{split}
\end{align}
One can show analytically that
\begin{equation}
\label{eqn:dtilde_det}
\det \big( \widetilde{D} \big) = \left( \left( c + d \right)^{L_{s}} + m_{f} \left( c - d \right)^{L_{s}} \right)^{12V}
\end{equation}
where $m_{f}$ is the fermion mass, $V$ is the 4D lattice volume, and $L_{s}$ is the number of $s$ sites. Since this has no dependence on the gauge field, $D_{\rm DWF}$ can be replaced with $D_{\rm EOFA}$ in the path integral without modifying the physics. This formalism has the advantage that $H \equiv \gamma_{5} R_{5} D_{\rm EOFA}$ is Hermitian even for M\"{o}bius DWF
, but comes at the cost of evaluating the dense 5D operator $D^{\perp}_{ss'}$. \\

The authors then show, using block manipulations in spin space and the Schur determinant identity, that the factorization \eqref{eqn:eofa_det_ratio} holds with $D = D_{\rm EOFA}$. Defining $\Delta_{\pm} \equiv R_{5} \left( D^{\perp}_{\pm}(m_{2}) - D^{\perp}_{\pm}(m_{1}) \right)$, and observing that $\Delta_{\pm}$ factorizes as $\Delta_{\pm} = k \Omega_{\pm} \Omega_{\pm}^{\dagger}$, the authors further demonstrate that this same determinant ratio can be written as a pseudofermion path integral in terms of the action $S_{f} = \phi^{\dagger} \mathcal{M}_{\rm EOFA} \phi$, with
\begin{equation}
\label{eqn:eofa_action}
\mathcal{M}_{\rm EOFA} = \mathbbm{1} - k P_{-} \Omega_{-}^{\dagger} \left[ H(m_{1}) \right]^{-1} \Omega_{-} P_{-} + k P_{+} \Omega_{+}^{\dagger} \left[ H(m_{2}) - \Delta_{+} P_{+} \right]^{-1} \Omega_{+} P_{+}.
\end{equation}
This is the final form of the EOFA action explored in this work. \\

In the following sections we perform tests of EOFA using two $N_{f} = 2+1$ RBC/UKQCD domain wall fermion ensembles. The properties of these ensembles are summarized in Table \ref{tab:ensembles}.
\begin{table}[h]
\centering
\resizebox{\linewidth}{!}{
\begin{tabular}{cccccccc}
\hline
\hline
\rule{0cm}{0.4cm}Ensemble & Action & $\beta$ & $L^{3} \times T \times L_{s}$ & M\"{o}bius scale & $a m_{l}$ & $a m_{h}$ & $m_{\pi}$ (MeV) \\
\hline
\rule{0cm}{0.4cm}16I \cite{Allton:2007hx} & DWF + I & 2.13 & $16^{3} \times 32 \times 16$ & --- & 0.01 & 0.032 & 400(11) \\
24ID \cite{course_ensembles} & MDWF + ID & 1.633 & $24^{3} \times 64 \times 24$ & 4.0 & 0.00107 & 0.0850 & 137.1(5) \\
\hline
\hline
\end{tabular}
}
\caption{Summary of the ensembles used in this work. (M)DWF denotes (M\"{o}bius) domain wall fermions, and I(D) denotes the Iwasaki gauge action (+ DSDR term) with coupling $\beta$.}
\label{tab:ensembles}
\end{table}

\vspace{-0.5cm}
\section{Hybrid Monte Carlo with EOFA}

The HMC algorithm generates a Markov chain of gauge field configurations by evolving a Hamiltonian system describing the coupled dynamics of the gauge field and fermions in (unphysical) molecular dynamics ``time''. In the following subsections we discuss the details of HMC for EOFA and our tests of each component.

\subsection{Action}

The EOFA action is given by \eqref{eqn:eofa_action}. We verify our implementation and the correctness of the formal derivation in Ref.~\cite{Chen:2014hyy} through the relationship suggested by \eqref{eqn:eofa_det_ratio} and \eqref{eqn:rhmc_det_ratio}: the EOFA and RHMC actions should compute the same determinant ratio up to the normalization factor \eqref{eqn:dtilde_det}. We verify that this is indeed true on a single configuration of the 16I and 24ID ensembles.

\begin{table}[h]
\centering
\begin{tabular}{cccccc}
\hline
\hline
\rule{0cm}{0.4cm}Ensemble & $N_{\rm hits}$ & $a m_{1}$ & $a m_{2}$ & RHMC & EOFA \\ 
\hline
\rule{0cm}{0.4cm}16I & 10 & 0.032 & 0.042 & 67.2(9) & 67.1(1) \\
24ID & 10 & 0.085 & 0.09 & 521.9(2.0) & 520.0(2) \\
\hline
\hline
\end{tabular}
\caption{Stochastic evaluations of $-\operatorname{logdet}(D_{\rm DWF}(m_{1})/D_{\rm DWF}(m_{2}))$ using RHMC and EOFA.}
\end{table}

\subsection{Heatbath}

At the start of each HMC trajectory we draw a random pseudofermion field $\phi$ according to $P(\phi) \propto \exp(-\phi^{\dagger} \mathcal{M}_{\rm EOFA} \phi)$. This is accomplished by generating a random Gaussian vector $\eta$, and then computing $\phi = \mathcal{M}_{\rm EOFA}^{-1/2} \eta$ using a rational approximation $x^{-1/2} \simeq \alpha_{0} + \sum_{l=1}^{N_{p}} \alpha_{l} / (\beta_{l} + x)$. Defining $\gamma_{l} = \left( 1 + \beta_{l} \right)^{-1}$, one can show that the resulting rational approximation to $\mathcal{M}_{\rm EOFA}^{-1/2}$ takes the form
\begin{equation}
\label{eqn:eofa_heatbath}
\resizebox{0.91\linewidth}{!}{
	$ \displaystyle \mathcal{M}_{\rm EOFA}^{-1/2} \simeq \alpha_{0} \mathbbm{1} + \sum_{l=1}^{N_{p}} \alpha_{l} \gamma_{l} \left\{ \mathbbm{1} + k \gamma_{l} P_{-} \Omega_{-}^{\dagger} \left[ H(m_{1}) - \gamma_{l} \Delta_{-} P_{-} \right]^{-1} \Omega_{-} P_{-} - k \gamma_{l} P_{+} \Omega_{+}^{\dagger} \left[ H(m_{2}) - \gamma_{l} \beta_{l} \Delta_{+} P_{+} \right]^{-1} \Omega_{+} P_{+} \right\} $
},
\end{equation}
requiring $2 N_{p}$ CG inversions to compute $\phi$. These inversions are not amenable to a multishift CG algorithm since the operators $\Delta_{\pm} P_{\pm}$ are singular, making the EOFA heatbath more expensive than the RHMC heatbath. In practice, we observe that the eigenvalues of $\mathcal{M}_{\rm EOFA}$ cover a relatively small interval, allowing us to accurately compute $\phi$ using a rational approximation with a modest number of poles and partially ameliorate this cost.

\subsection{Pseudofermion Force}

\begin{figure}[h]
\centering
\subfloat{\includegraphics[width=0.31\linewidth]{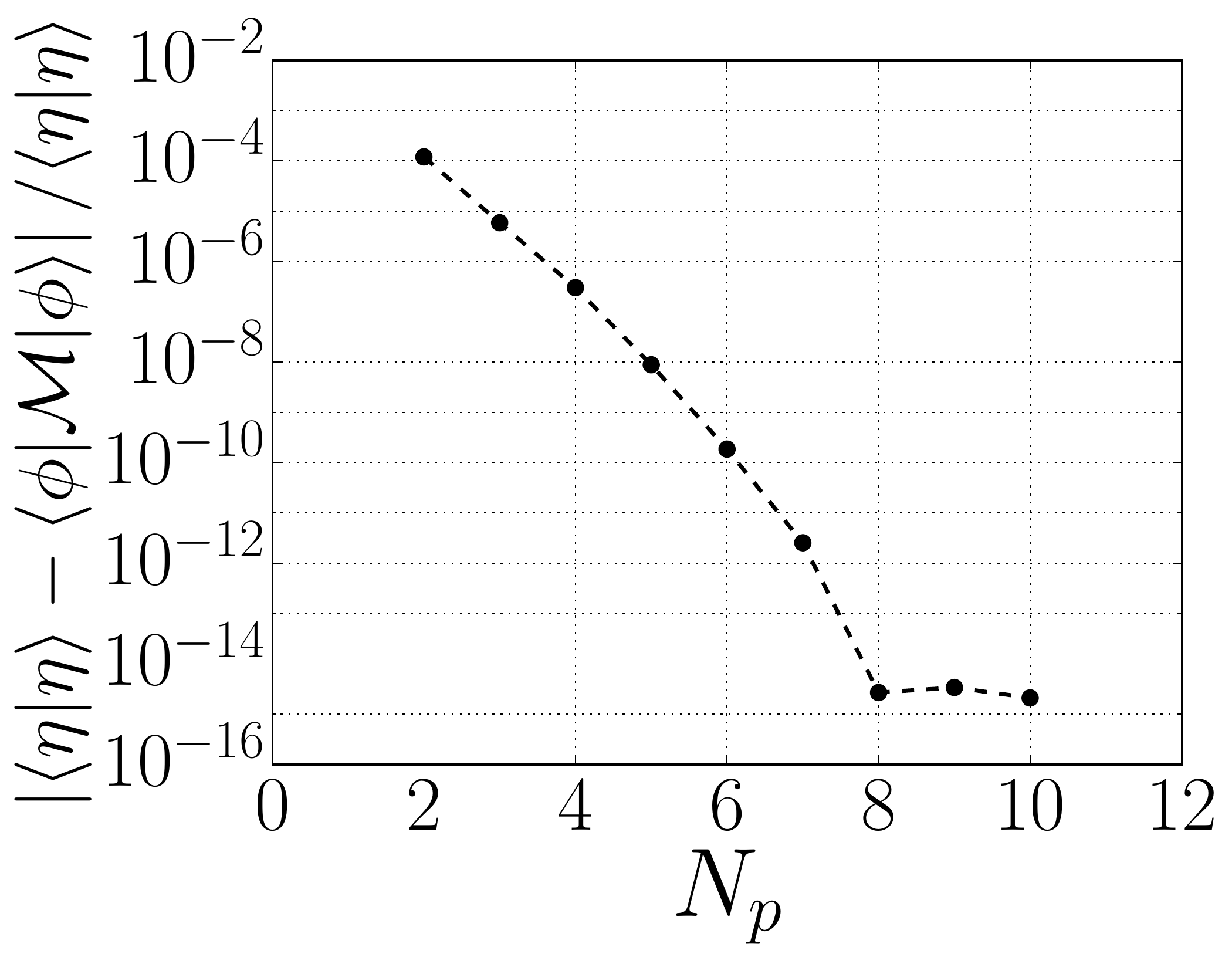}} \quad
\subfloat{\includegraphics[width=0.31\linewidth]{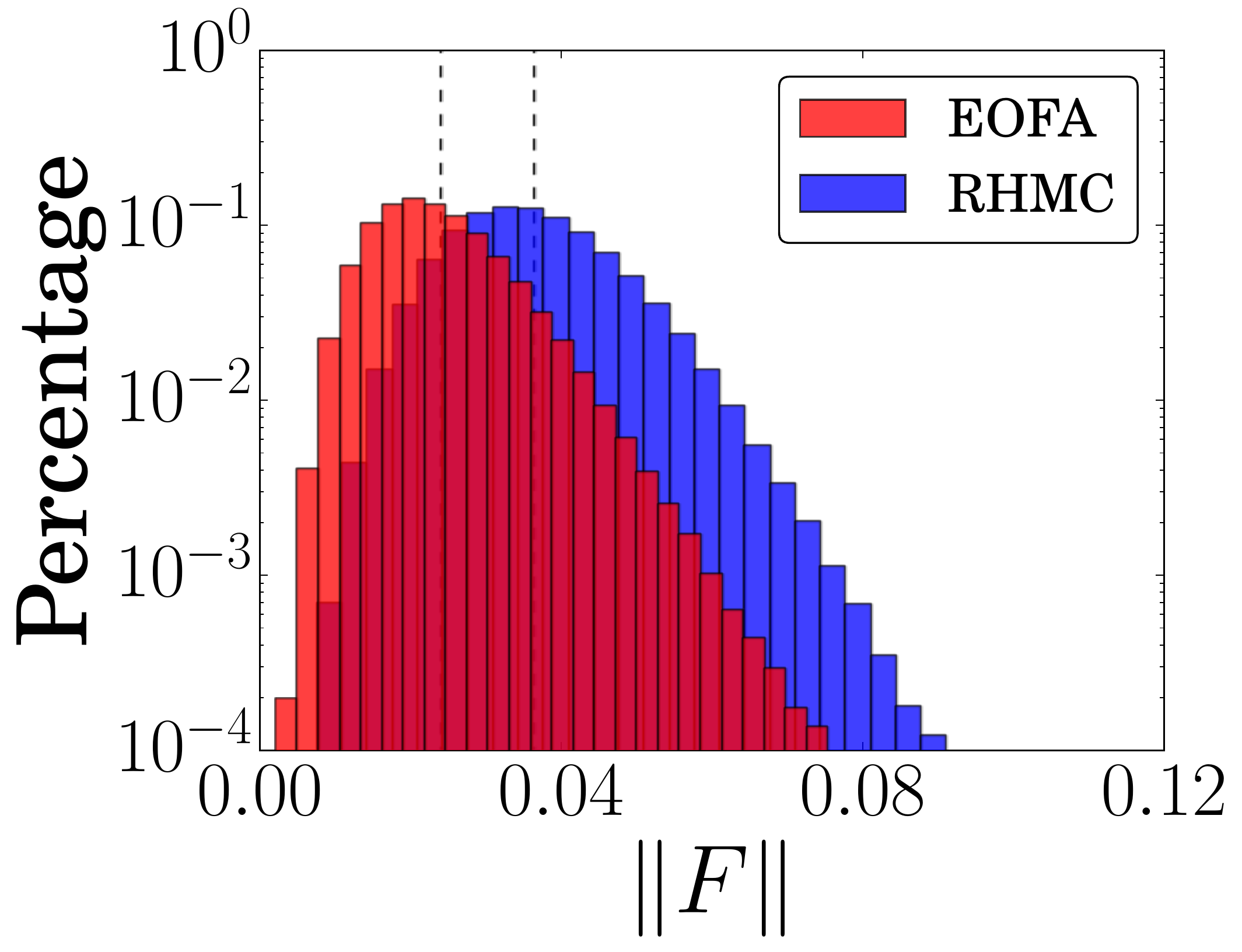}} \quad
\subfloat{\includegraphics[width=0.31\linewidth]{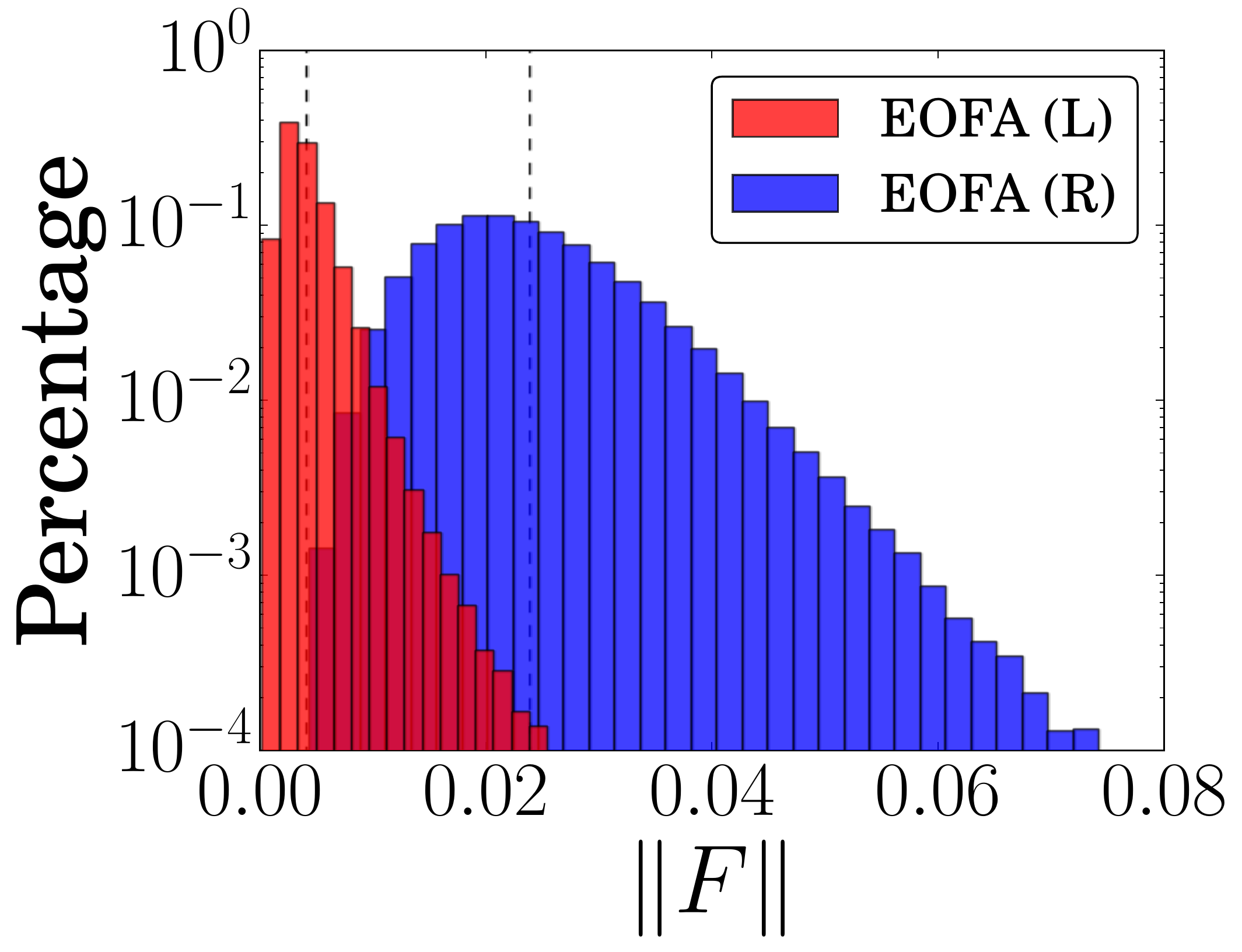}}
\caption{Left (24ID): Relative error in the heatbath step as a function of the number of poles in the rational approximation to $\mathcal{M}_{\rm EOFA}$. Middle (16I): Lattice-wide distribution of EOFA and RHMC total forces by gauge link. Right (16I): Lattice-wide distribution of EOFA force contributions from L and R terms by gauge link. Vertical dashed lines mark the average of each distribution.}
\label{fig:16I_heatbath_force}
\end{figure}

The EOFA pseudofermion force is derived by varying \eqref{eqn:eofa_action} with respect to the gauge field:
\begin{equation}
\label{eqn:eofa_force}
\partial_{x,\mu}^{a} S_{f}[U] = k \chi_{L}^{\dagger} \gamma_{5} R_{5} \left( \partial_{x,\mu}^{a} D_{w} \right) \chi_{L} - k \chi_{R}^{\dagger} \gamma_{5} R_{5} \left( \partial_{x,\mu}^{a} D_{w} \right) \chi_{R},
\end{equation}
with $\chi_{L} \equiv \left[ H(m_{1}) \right]^{-1} \Omega_{-} P_{-} \phi$ and $\chi_{R} \equiv \left[ H(m_{2}) - \Delta_{+} P_{+} \right]^{-1} \Omega_{+} P_{+} \phi$. This can be evaluated at the cost of two CG inversions, in contrast to the corresponding RHMC force evaluations, which require three multishift CG inversions. In Figure \ref{fig:16I_heatbath_force} we plot distributions of the magnitude of the pseudofermion force associated with each gauge link on a single configuration of the 16I ensemble, and confirm TWQCD's observation that the average EOFA force is somewhat smaller in magnitude than the average RHMC force.
While TWQCD has reported a speed-up by using a Sexton-Weingarten integration scheme to exploit the asymmetry in the size of the left-handed and right-handed EOFA force contributions \cite{Chen:2014bbc}, we have yet to explore this direction in our work.

\section{Reproduction of the 16I Ensemble Using EOFA}

As a final test of EOFA we ran two, parallel streams to reproduce the 16I ensemble using the parameters in Ref.~\cite{Allton:2007hx}. On one stream the RHMC action was used to evolve the strange quark, and on the other stream the EOFA action was used to evolve the strange quark; otherwise the evolutions are identical. We generated 1500 trajectories of each stream, and then compared a number of observables --- including the plaquette, topological susceptibility, and pion and kaon masses and decay constants --- measured every ten trajectories from 500 to 1500. We find complete agreement within the computed statistical errors. We also observe similar rates of topological tunneling, with the caveat that a proper study of the autocorrelation time would require a longer run. 
\begin{table}[!ht]
\vspace{-0.24cm}
\centering
\begin{tabular}{ccc|ccc}
\hline
\hline
\rule{0cm}{0.3cm}& EOFA & RHMC & & EOFA & RHMC\\
\hline
\rule{0cm}{0.3cm}RMS $\left< \Delta H \right>$ & 0.23(2) & 0.20(2) & $a m_{\pi}$ & 0.243(2) & 0.244(1) \\
$\left< \exp(-\Delta H) \right>$ & 1.00(2) & 0.98(2) & $a m_{K}$ & 0.326(2) & 0.326(1) \\
$\left< P \right>$ & 0.58807(3) & 0.58808(4) & $a m_{\Omega}$ & 0.990(9) & 0.995(10) \\
$\chi_{t} = \left<Q^{2}\right> / V$ & $4.5(7) \times 10^{-5}$ & $3.5(5) \times 10^{-5}$ & $a f_{\pi}$ & 0.0887(8) & 0.0883(6) \\
$a m_{\rm res}'(m_{l})$ & 0.00305(4) & 0.00306(4) & $a f_{K}$ & 0.0966(6) & 0.0963(4) \\
\hline
\hline
\end{tabular}
\caption{Results for some basic observables on the 16I ensemble in lattice units.}
\label{tab:16I_reproduction}
\end{table}

\vspace{-0.84cm}
\section{Performance Improvements}
\label{sec:optimizations}

In this section we discuss various algorithmic refinements to the basic EOFA formalism, and benchmark EOFA against RHMC on the 24ID ensemble. Each MD trajectory of the 24ID ensemble consists of 12 steps of a nested force gradient QPQPQ integrator, with the strange quark on the outermost time step. Timings are reported for a 256-node BG/Q partition.

\vspace{-0.15cm}
\subsection{Accelerating EOFA Inversions}
\label{sec:inversion_optimizations}

\begin{figure}[!ht]
\centering
\subfloat{\includegraphics[width=0.45\linewidth]{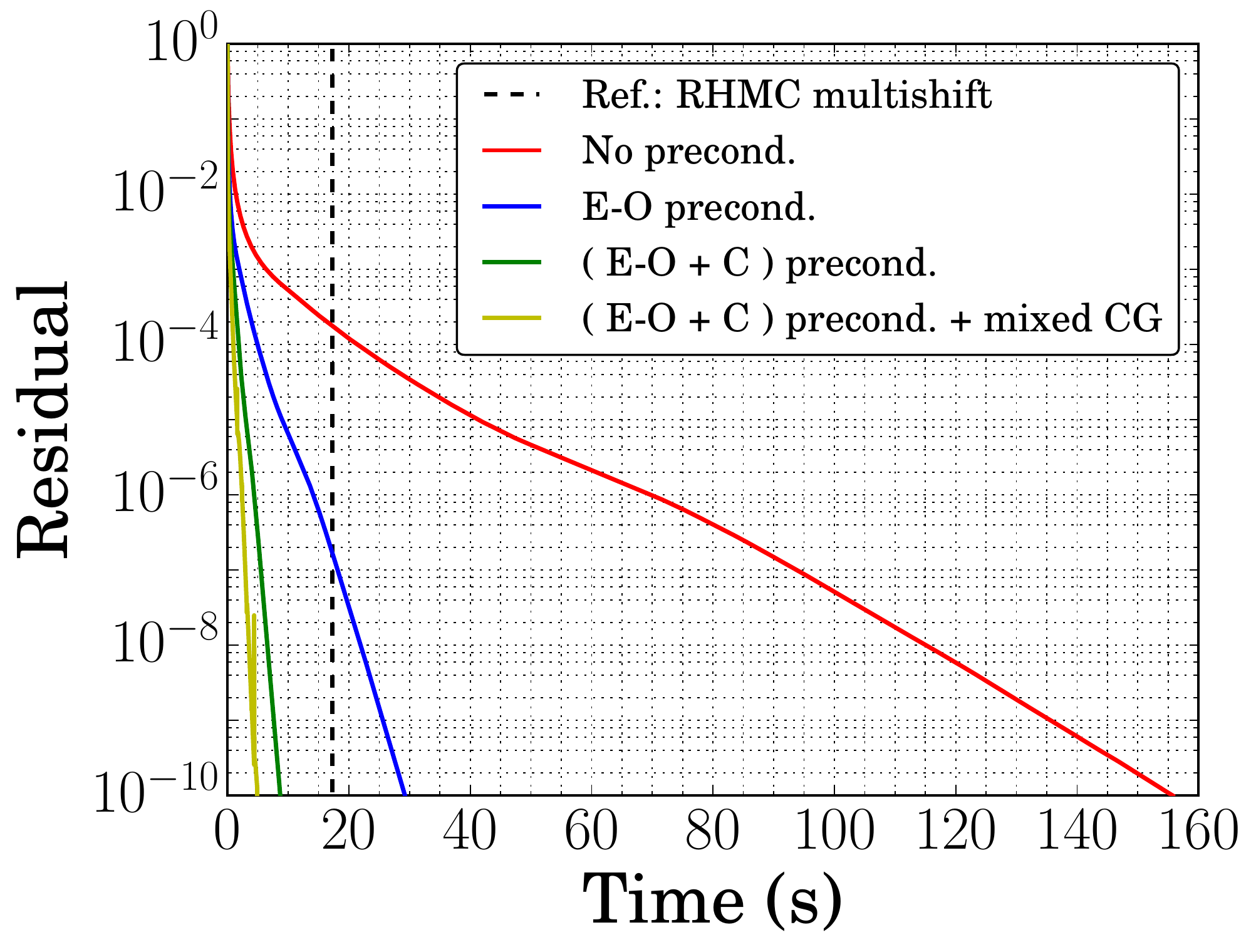}} \quad
\subfloat{\includegraphics[width=0.45\linewidth]{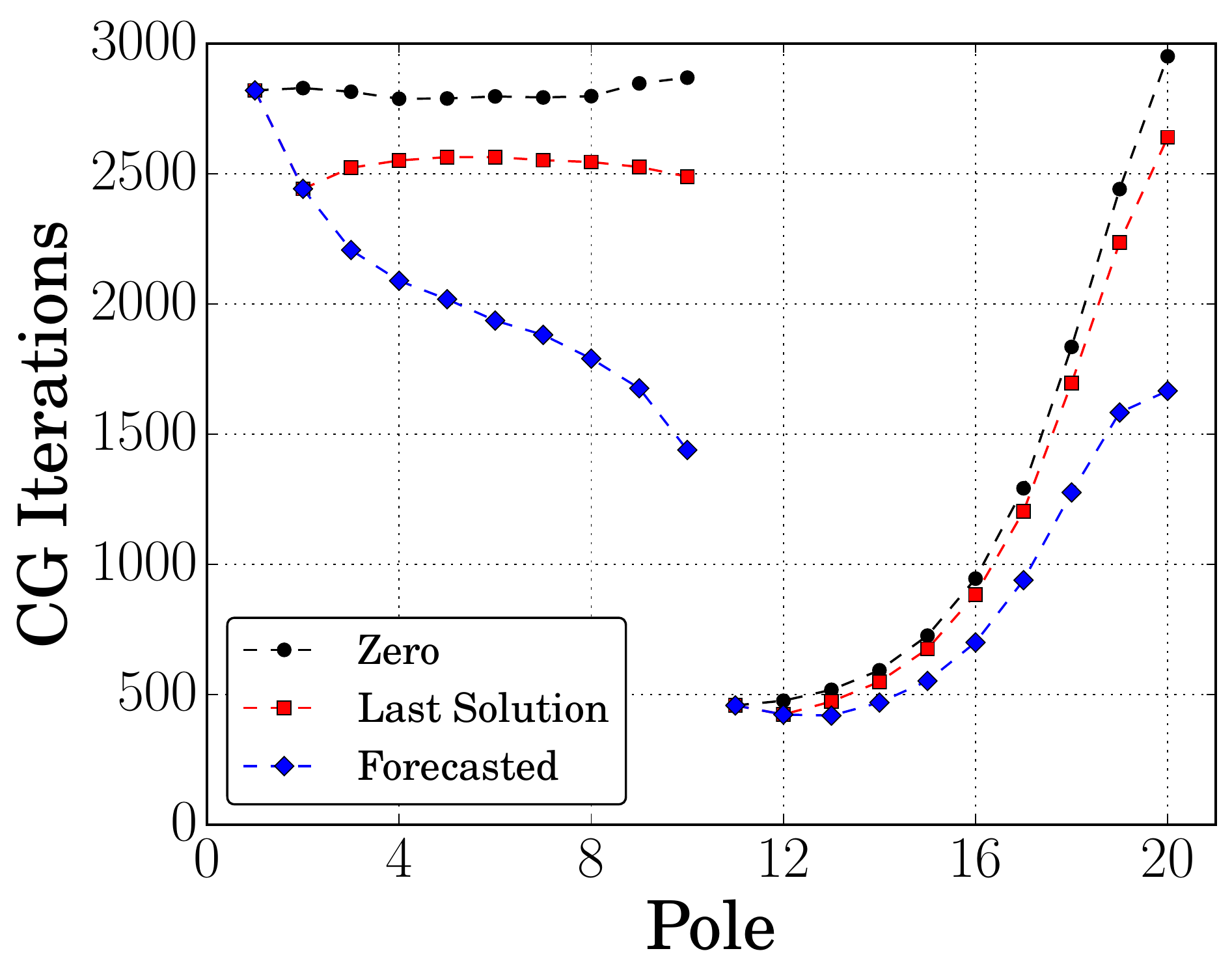}}
\caption{Left: wall clock inversion time for a single EOFA strange quark solve on the 24ID ensemble as the acceleration techniques described in the text are introduced. We find an overall $31.8 \times$ speed-up, and a $3.5 \times$ speed-up relative to an even-odd preconditioned multishift inversion of $D_{\rm DWF}^{1/2} \psi = \phi$ at the strange quark mass. Right: CG iteration counts for each solve in the EOFA heatbath comparing different schemes for the initial guesses.}
\label{fig:optimizations}
\end{figure}

Since the majority of the computational effort in an HMC evolution lies in repeatedly inverting the Dirac operator, techniques to accelerate these inversions can lead to substantial increases in the overall efficiency of the evolution. In the context of EOFA, the linear system we invert takes the general form
\begin{equation}
\label{eqn:eofa_linear_system}
\left( H + \alpha_{l} \Delta_{\pm} P_{\pm} \right) \psi = \phi.
\end{equation}
We have introduced a number of such refinements, including: even-odd preconditioning, Cayley-form preconditioning\footnote{This is a novel technique specific to EOFA, which exploits the relationship $D_{\rm DWF} = D_{\rm EOFA} \cdot \widetilde{D}$ from Eqn.~\eqref{eqn:Deofa_def} to right-precondition Eqn.~\eqref{eqn:eofa_linear_system}, resulting in an equivalent linear system in terms of $D_{\rm DWF}$ rather than $D_{\rm EOFA}$. Since the 5D structure of $D_{\rm EOFA}$ is dense, whereas the 5D structure of $D_{\rm DWF}$ has a tridiagonal Cayley form, this results in a preconditioned system that is substantially cheaper to solve.}, and the use of mixed-precision CG.
In the left panel of Figure \ref{fig:optimizations} we show the successive improvements in inversion time as each of these techniques is introduced for a single inversion of \eqref{eqn:eofa_linear_system} at the strange quark mass on the 24ID ensemble.

\subsection{Forecasted Solutions for the Heatbath Step}
\label{sec:heatbath_optimizations}

The EOFA heatbath requires \eqref{eqn:eofa_linear_system} to be simultaneously solved for $2 N_{p}$ values of $\alpha_{l}$, arising from the rational approximation to $\mathcal{M}_{\rm EOFA}^{-1/2}$. Like TWQCD, we use the chronological inversion technique introduced by Brower et al.~\cite{Brower:1995vx} to forecast solutions for a given $\alpha_{l}$ from previous solutions for other $\{ \alpha_{l} \}$. We observe that by the tenth pole the iteration count has been approximately halved relative to using zero or the solution for the previous $\alpha_{l}$ as the initial CG guess.

\subsection{HMC Timing Benchmarks}

In Table \ref{tab:hmc_timing_benchmark} we compare timings for a single MD trajectory using RHMC and EOFA with and without Cayley-form preconditioning. The details of the ensemble parameters and force gradient integrator are identical except for the choice of strange quark action. We find that without this technique RHMC and EOFA break even: the expensive heatbath and cost of inverting $D_{\rm EOFA}$ negate the expected gain from the simpler form of the energy and force evaluations. Once Cayley-form preconditioning is introduced, we observe a $2.4 \times$ speed-up over RHMC. We expect that by optimizing our code and retuning the details of the force gradient integrator for EOFA a speed-up of $3 \times$ or more should be possible.
\begin{table}[!ht]
\centering
\resizebox{0.95\linewidth}{!}{
\begin{tabular}{c|cc|cc|cc}
\hline
\hline
& \multicolumn{2}{|c|}{\textbf{RHMC}} & \multicolumn{2}{|c|}{\textbf{EOFA} (dense)} & \multicolumn{2}{|c}{\textbf{EOFA} (Cayley precond.)} \\
Step & Time (s) & \% & Time (s) & \% & Time (s) & \% \\
\hline
\rule{0cm}{0.4cm}Heatbath & 42.9 & 2.0 & 340.6 & 15.1 & 160.1 & 18.4 \\
Force gradient integration (total) & 1865.2 & 88.9 & 1840.6 & 81.8 & 684.0 & 78.7 \\
Final Hamiltonian evaluation & 189.4 & 9.0 & 68.8 & 3.0 & 25.0 & 2.9 \\
\hline
\rule{0cm}{0.4cm}Total & 2097.5 & --- & 2250.0 & --- & 869.1 & --- \\
(Total RHMC) / Total & 1.00 & --- & 0.93 & --- & 2.41 & --- \\
\hline
\hline
\end{tabular}
}
\caption{Strange quark timings for a single 24ID MD trajectory on a 256-node BG/Q partition.}
\label{tab:hmc_timing_benchmark}
\end{table}

\section{Conclusion}
We have independently implemented and tested TWQCD's exact one-flavor algorithm. We find, after optimizing, that the HMC evolution of the strange quark is 2.4 times faster per trajectory with EOFA for a $24^{3} \times 64 \times 24$ physical mass M\"{o}bius domain wall fermion ensemble. The key to this improvement is a preconditioning technique that relates inversions of $D_{\rm EOFA}$ to cheaper inversions of $D_{\rm DWF}$. We expect that further improvements are possible, and are working to implement EOFA with G-parity boundary conditions for our ongoing $I=0$ $K \rightarrow \pi \pi$ calculation \cite{g-parity}. We will elaborate on the details of our EOFA implementation in a forthcoming publication \cite{EOFA_paper}.

\section{Acknowledgments}

The authors would like to thank members of the RBC/UKQCD collaboration for helpful discussions. This work was supported in part by U.S. DOE grant \#DE-SC0011941. Calculations were performed on the BG/Q computers at Brookhaven National Lab.



\begin{thebibliography}{99}
\bibitem{Hasenbusch:2001} M.~Hasenbusch, ``Speeding up the Hybrid Monte Carlo Algorithm for Dynamical Fermions'', Phys. Lett. \textbf{B}519, 177-182 (2001).
\bibitem{zmobius} T.~Blum et al., ``zM\"{o}bius and Other Recent Developments on Domain Wall Fermions'', Talk Presented at the 33rd International Symposium on Lattice Field Theory, Kobe, Japan (2015).
\bibitem{greg_thesis} G.~McGlynn, ``Advances in Lattice Quantum Chromodynamics'', Ph.D. Thesis, Columbia University (2016).
\bibitem{Chen:2014hyy} Y.C.~Chen and T.W.~Chiu, ``Exact Pseudofermion Action for Monte Carlo Simulation of Domain-Wall Fermion'', Phys. Lett. \textbf{B}738, 55-60 (2014).
\bibitem{g-parity} Z. Bai et al., ``Standard Model Prediction for Direct $CP$ Violation in $K \rightarrow \pi \pi$ Decay'', Phys. Rev. Lett. \textbf{115}, 212001 (2015).
\bibitem{Ogawa:2009ex} K.~Ogawa, T.W.~Chiu, and T.H.~Hsieh, ``One-Flavor Algorithm for Wilson and Domain-Wall Fermions'', PoS \textbf{LATTICE2009}, 033 (2009).
\bibitem{Chen:2014bbc} Y.C.~Chen and T.W.~Chiu, ``One-Flavor Algorithms for Simulation of Lattice QCD with Domain-Wall Fermion: EOFA versus RHMC'', PoS \textbf{LATTICE2014}, 059 (2014). 
\bibitem{Allton:2007hx} C.~Allton et al., ``2+1 Flavor Domain Wall QCD on a $(2 \,\, \mathrm{fm})^3$ Lattice: Light Meson Spectroscopy with $L_{s} = 16$'', Phys. Rev. \textbf{D}76, 014504 (2007).
\bibitem{course_ensembles} R.D.~Mawhinney, ``Scaling and Properties of $1/a = 1$ GeV, 2+1 Flavor M\"{o}bius Domain Wall Fermion Ensembles'', to appear in PoS \textbf{LATTICE2016}.
\bibitem{Brower:1995vx} R.C.~Brower et al., ``Chronological Inversion Method for the Dirac Matrix in Hybrid Monte Carlo'', Nucl. Phys. \textbf{B}484, 353-374 (1997).
\bibitem{EOFA_paper} D.J.~Murphy et al., ``Accelerating M\"{o}bius Domain Wall Fermion Simulations with the Exact One-Flavor Algorithm'', in preparation.	
\end{thebibliography}
\end{document}